\documentclass[superscriptaddress,twocolumn,a4paper,ajp]{revtex4}

\usepackage{graphicx}
\usepackage{fancyhdr}
\usepackage{amsmath}
\usepackage{amsfonts}

\newcommand{\ndpar}[3] { \frac{\partial^{#3} {#1} } {\partial #2 ^{#3} }}

\newcommand{\OL}[1] {\textrm{\emph{\large O}}\!\left({#1}\right)}
\newcommand{\tr}[0] {\mbox{tr}\,}

\newcommand{\II}[0] {\mbox{i}}

\newcommand{\dis}[0] {\displaystyle}

\newcommand{\abs}[1]{\left| #1 \right|}

\begin{document}

\title{ Tunnelling in quantum superlattices with variable
lacunarity}

\author{Francisco R. Villatoro}
 \affiliation{Departamento de Lenguajes y Ciencias de la Computaci\'on,
  Universidad de M\'alaga, E-29071 M\'alaga, Spain}

\author{Juan A. Monsoriu}
 \email[Corresponding author:]{jmonsori@fis.upv.es}
 \affiliation{Departamento de F{\'\i}sica Aplicada,
 Universidad Polit\'ecnica de Valencia, E-46022 Valencia, Spain}



\date{\today}

\begin{abstract}

Quantum fractal superlattices are microelectronic devices consisting
of a series of thin layers of two semiconductor materials deposited
alternately on each other over a substrate following the rules of
construction of a fractal set, here, a symmetrical polyadic Cantor
fractal. The scattering properties of electrons in these
superlattices may be modeled by using that of quantum particles in
piecewise constant potential wells. The twist plots representing the
reflection coefficient as function of the lacunarity parameter show
the appearance of black curves with perfectly transparent tunnelling
which may be classified as vertical, arc, and striation nulls.
Approximate analytical formulae for these reflection-less curves are
derived using the transfer matrix method. Comparison with the
numerical results show their good accuracy.
\end{abstract}

\maketitle

\section{Introduction}
\label{Intro}

Quantum superlattices are microelectronic devices composed of
alternating layers of two semiconductors with different energy band
gaps deposited over a
substrate~\cite{BOOKQuantumWell,BOOOKHeterostructures,QuantumWells,Superlattices}.
The sandwiching of two materials whose band gap energies are
different results in a retangular potential well for the electrons
(holes) in the conduction (valence) band. The thickness of each
semiconductor layer is about 10--100 atoms (1--10 nm), so this
three-dimensional finite potential well may be split into a
two-dimensional system in the transversal dimensions, where
electrons behave as a two-dimensional electron gas with continuous
energy states, and a one-dimensional finite square well in the
growth direction with discrete bound states.

In the absence of external electric field, semiconductor
superlattices may be considered, in the first approximation, as
quasi-one-dimensional system of rectangular quantum wells separated
by potential barriers (in the conduction band). When a external
force is applied to the electrons in the quantum-well
heterostructure they do not respond as free particles due to the
influence of the other atoms. These effects may be taken into
account by introducing an ``effective mass" ($m^*$) which relates
the particle motion to an external force (or potential) without
worrying about all the atomic forces. Therefore, the scattering in
semiconductor superlattices can be easily solved by means of the
transfer matrix method using an effective mass for the
electrons~\cite{NawrockiAJP,KalotasEJP,SprungAJP,SprungAJP2,WalkerAJP}.

Fractal superlattices are designed by alternating semiconductor
materials according to an iterative fractal process using techniques
such as molecular beam epitaxy~\cite{BookMBE}. These devices may be
crystalline or amorphous. A typical device for the first kind is
made by several layers of gallium arsenide (GaAs) of a few Angstrom
(\AA) thick sandwiched between aluminium gallium arsenide
(Al$_x$Ga$_{1-x}$As) ones, where the crystal composition $x$ may
vary between 0 and 1, such as those fabricated by Axel and
Terauchi~\cite{Axel91} using $x=0.3$, corresponding to a monolayer
of Al$_x$Ga$_{1-x}$As with thickness of 0.28 nm. A typical amorphous
device is made by alternating layers of amorphous germanium (a-Ge)
and amorphous silicon (a-Si) deposited in a silicon
substrate~\cite{ExpCantorSuperlattices}, each layer having a
thickness of less than 1.4 nm with good sharpness.

The scattering of electrons in symmetrical polyadic Cantor fractal
potentials, characterized by a lacunarity parameter which is
independent of their fractal dimension, have been numerically
obtained by means of the transfer matrix method in
Ref.~\cite{VillatoroAJP}. The representation of the reflection
coefficient as a function of the particle energy and the lacunarity
shows perfectly transparent tunneling along some curves, referred to
as nulls, which may be classified as vertical
(lacunarity-independent), arc (concave upward), and striation
(concave downward) nulls~\cite{JaggardJOSA}. Analytical expressions
for the position of these nulls have been obtained in
Ref.~\cite{JaggardJOSA} in an optical context, which may be properly
adapted to the present application. However, those authors omit the
details of their derivation which may be of great interest when
quantum fractal superlattices are incorporated as laboratory
assignments~\cite{VillatoroAJP,VillatoroEJP}.

In this paper, the transfer matrix method is used to obtain
analytical expressions for the reflection-less energies of particles
in fractal superlattices designed as polyadic Cantor sets with
variable lacunarity. The contents of this paper are as follows. The
next section recalls the main facts about these kind of fractal
superlatices. Section~3, presents the analytical formulas for the
reflection-less energies, whose validity is shown in Section~4 by
its comparison with numerical results. Finally, the last section is
devoted to the main conclusions.

\section{Polyadic Cantor superlattices}
\label{Superlattices}

\begin{figure}
 \begin{center}
  \includegraphics[width=8cm]{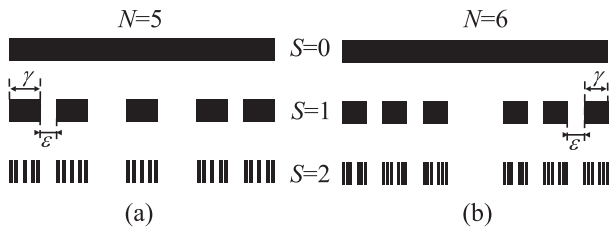}
 \end{center}
  \caption{First two stages ($S=1$ and $S=2$) of
  polyadic Cantor fractal sets with $N=5$ (a) and $N=6$ (b),
  showing the definition
  of both the scale factor ($\gamma$) and the lacunarity parameter
  ($\varepsilon$). Black and white regions denote the potential
  values $-V$ and $0$, respectively.}
 \label{PolyadicCantorSet}
\end{figure}

In Cantor superlattices the distribution of the layers' thickness
along growth axis corresponds to a finite order approximation to a
polyadic, or generalized, Cantor set~\cite{Fractals,Mandelbrot}.
These sets are defined as follows. The first step ($S=0$) is to take
a segment of unit length, referred to as initiator. In the next
step, $S=1$, the segment is replaced by $N$ non-overlapping copies
of the initiator, each one scaled by a factor $\gamma<1$. For even
$N$, as shown in Fig.~\ref{PolyadicCantorSet}(b), one half of the
copies are placed completely to the left of the interval and the
other half completely to its right, each copy being separated by a
fixed distance, let us say $\varepsilon$. For odd $N$, as shown in
Fig.~\ref{PolyadicCantorSet}(a), one copy lies exactly centered in
the interval, and the rest ones are distributed as for even $N$,
i.e., $\lfloor N/2 \rfloor$ copies are placed completely to the left
of the interval and the other $\lfloor N/2 \rfloor$ copies wholly to
its right, where $\lfloor N/2 \rfloor$ is the greatest integer less
than or equal to $N/2$. At the following construction stages of the
the fractal set, $S=2, 3, \ldots$, the generation process is
repeated over and over again for each segment in the previous stage.
Strictly speaking the Cantor set is the limit of this procedure for
$S\rightarrow\infty$, which is composed of geometric points
distributed such that each point lies arbitrarily close of other
points of the set, being the $S$-th stage Cantor set usually
referred to as a pre-fractal or physical fractal.

Symmetrical polyadic Cantor fractals are characterized by three
parameters, the number of self-similar copies $N$, the scaling
factor $\gamma$, and the width of the outermost gap at the first
stage, $\varepsilon$, here on referred to as the lacunarity
parameter, as in most of the previous technical papers dealing with
polyadic Cantor fractals in engineering
applications~\cite{JaggardJOSA,MonsoriuOE}. The similarity dimension
of all polyadic Cantor fractals is
$D=\ln(N)/\ln(\gamma^{-1})=-\ln(N)/\ln(\gamma)$, which is
independent of the lacunarity parameter. The three parameters of a
polyadic Cantor set must satisfy certain constraints in order to
avoid overlapping between the copies of the initiator. On the one
hand, the maximum value of the scaling factor depends on the value
of $N$, such that $0<\gamma <\gamma_{\mbox{\scriptsize max}}= 1/N$.
On the other hand, for each value of $N$ and $\gamma$, there are two
extreme values for $\varepsilon$. The first one is
$\varepsilon_{\mbox{\scriptsize min}}=0$, for which the highest
lacunar fractal is obtained, i.e., that with the largest possible
gap. For even $N$, the central gap has a width of $1-N\,\gamma$, and
for odd $N$, both large gaps surrounding the central well have a
width of $(1-N\,\gamma)/2$. The other extreme value is
\[
 \varepsilon_{\mbox{\scriptsize max}}=
 \left\{ \begin{array}{ll}
 \dis \frac{1-N\,\gamma}{N-2},\qquad & \mbox{even } N, \\
  \\
 \dis \frac{1-N\,\gamma}{N-3},\qquad & \mbox{odd } N,
 \end{array}\right.
\]
where for even (odd) $N$ two (three) wells are joined together in
the center, without any gap in the central region. The width of the
$N-2$ gaps in this case is equal to $\varepsilon_{\mbox{\scriptsize
max}}$. Thus the corresponding lacunarity is lower than that for
$\varepsilon=0$, but not the smallest one, which is obtained for the
most regular distribution, where the gaps and wells have the same
width at the first stage ($S=1$) given by
\[
 \varepsilon_{\mbox{\scriptsize reg}}= \frac{1-N\,\gamma}{N-1}.
\]
Note that $0 < \varepsilon_{\mbox{\scriptsize reg}} <
\varepsilon_{\mbox{\scriptsize max}}$.

Polyadic Cantor superlattices have been experimentally grown by
J\"arrendahl et al.~\cite{ExpCantorSuperlattices} by means of
properly alternating layers of a-Ge and a-Si. X-ray diffraction
techniques were used to analyze the scattering problem of electrons
in structures such a $S=5$ triadic Cantor pre-fractal having $3^6$
(729) ``elementary" alternating layers. A theoretical analysis of
the diffraction peaks has shown good agreement with the results of
the experiment. However, a complete analysis requires taking into
account both sample imperfections and instrumental limitations,
which can influence location, intensity, and profile of the peaks.
In this paper, in order to simplify the analysis, these effects are
neglected.

\section{Estimation of the null positions}

The graphical representation of the scattering reflection
coefficient as a function of both the particle energy and the
lacunarity parameter for polyadic Cantor pre-fractal potentials,
referred to as a twist plot, presents a characteristic null
structure (black lines and curves) corresponding to energies at
which the particle transparently tunnels through the fractal
potential wells, i.e., for which the reflection coefficient is null.

The reflection coefficient nulls may be separated in three main
families. The first and most noticeable is the family of vertical
nulls, being caused by the bound states of the particle in each
potential well, being thus independent of the lacunarity parameter.
The nulls in the second family, denoted as arc nulls, are concave
upward and curve from the upper left of the twist plots to the lower
right. The third family of nulls, referred to as striation nulls,
form the finer structure of the twist plots being curves which are
concave downward, running from the lower left to the upper right of
each twist plot.

The nulls may be calculated from first principles as previously done
in Ref.~\cite{JaggardJOSA} for the interference of light in
dielectric fractal superlattices, where only the final mathematical
expression is presented, completely omitting the details of the
derivation. In the next subsections this gap is filled for quantum
fractal superlattices.

\subsection{Transfer matrix method} \label{Problem}

The scattering problem for polyadic Cantor sets with variable
lacunarity can be easily solved by means of the transfer matrix
method using an effective mass for the electrons. Let us summarize
the main details of this widely known technique.

The quantum scattering of particles on one-dimensional potential
wells is governed by the steady-state, linear Schr\"odinger equation
\begin{equation}
 \label{LSE}
 -\frac{\hbar^2}{2\,m^*}\,\ndpar{\psi(x)}{x}{2} + V(x)\,\psi(x) =
 E\,\psi(x),
\end{equation}
where $\psi(x)$, $m^*$ and $E$ are the particle wavefunction,
effective mass and energy, respectively, $\hbar$ is Planck's
constant, and $V(x)$ describes a distribution of potential wells,
from here on, taken as a piecewise constant function. The
wavefunction $\psi_i$ on the region where the potential constant
value is $V_i$, is the addition of two plane waves,
$\psi_i(x)=\psi_i^{+}(x)+\psi_i^{-}(x)$, given by
\[
 \psi^\pm_i(x) = A^\pm_i\,e^{\dis  \pm\II\,k_i\,x}, \qquad
 k_i = \frac{1}{\hbar}\,\sqrt{2\,m^*\,(E-V_i)},
\]
where $\II=\sqrt{-1}$, $k_i$ is the local particle momentum, and
$A^\pm_i$ are integration constants to be determined by applying the
standard boundary conditions at the interfaces between successive
wells. In this paper, a distribution of potential wells of the same
depth is considered, so the potential $V_i=-V$ in the wells and
$V_i=0$ outside them.

The solution of Eq.~\eqref{LSE} for a distribution of $N$ constant
potential wells is easily obtained by means of the transfer matrix
method~\cite{KalotasEJP,SprungAJP,WalkerAJP,SprungAJP2,VillatoroEJP},
yielding
\begin{equation}
 \label{def:M}
 \left(\begin{array}{c}
    A^+_{0} \\ A^-_{0}
 \end{array}\right)
   = M\,
 \left(\begin{array}{c}
    A^+_{N+1} \\ A^-_{N+1}
 \end{array}\right),
\end{equation}
where
\[
 M = D_{0}^{-1}\,
   \left(
    \prod_{i=1}^N
       D_{i}\,P_i(d_i)\,D_{i}^{-1}
    \right)\,D_{N+1},
\]

\[
 D_i =
 \left(\begin{array}{cc}
    1 &  1 \\ k_i & -k_i
 \end{array}\right),
 \qquad
 P_i(d_i)
    =
 \left(\begin{array}{cc}
    e^{\dis  \II\,k_i\,d_i} & 0 \\
    0  & e^{\dis -\II\,k_i\,d_i}
 \end{array}\right),
\]
where $d_i$ is the width of the $i$-th potential well.

Both the reflection and transmission coefficients of the scattering
of a quantum particle, incoming from the left, with the $N$-well
potential is determined by the coefficients of the matrix $M$,
\[
 \left(\begin{array}{c}
    A^+_{0} \\ A^-_{0}
 \end{array}\right)
   = \left(\begin{array}{cc}
    M_{11} & M_{12} \\
\\
    M_{21} & M_{22}
 \end{array}\right) \,
 \left(\begin{array}{c}
    A^+_{N+1} \\ 0
 \end{array}\right),
\]
where no backward particle can be found on the right side of the
potential, so $A^-_{N+1}=0$. Both the reflection and transmission
coefficients~\cite{Liboff} are given by
\begin{equation} \label{reflection:transmission}
\begin{split}
 R &= \frac{\abs{A^-_{0}}^2}{\abs{A^+_{0}}^2} =
  \frac{\abs{M_{21}}^2}{\abs{M_{11}}^2},\\
\\
 T &= \frac{\abs{A^+_{N+1}}^2}{\abs{A^+_{0}}^2} =
  \frac{1}{\abs{M_{11}}^2},
\end{split}
\end{equation}

respectively, since $k_{N+1}=k_0$.

\subsection{Vertical nulls} \label{VerticalNulls}

The vertical nulls shown in the twist plots are the result of bound
states of the particle due to its resonance on every potential well
of the Cantor superlattice. Since there are $N^S$ potential wells at
the $S$-th stage pre-fractal, the vertical nulls are very noticeable
in the twist plot. Using the transfer matrix method is easy to
calculate their exact positions, as shown in the appendix of
Ref.~\cite{VillatoroAJP}. This derivation is repeated here for
completeness of the presentation.

Let us consider a quantum well at the $S$-th stage pre-fractal,
whose width is $a=L\,\gamma^S$, where $L$ is the length of the
initiator, separated from the next one by a distance
$b=L\,\epsilon^S=a\,(\epsilon/\gamma)^S$. For future convenience,
let us take a symmetrical configuration in which the quantum well is
surrounded by two regions of width $b/2$, which corresponds to
$V_0=V_2=0$, $V_1=-V$, $d_1=a$, and $d_0=d_2=b/2$, in the notation
introduced in Section~\ref{Problem}. The particle momentum at the
three constant potential regions may be normalized as

\begin{equation} \label{normalizationEnergy}
\begin{split}
 k_0&=k_2=\frac{\sqrt{2\,m^*\,E}}{\hbar} =\frac{\phi}{a},\\
\\
 k_1&=\frac{\sqrt{2\,m^*\,(E+V)}}{\hbar}
  =\frac{\sqrt{\phi^2+\phi_V^2}}{a},
\end{split}
\end{equation}

where $\phi$ and $\phi_V$ are the (non-dimensional) particle energy
and depth of the potential well, respectively. The total scattering
matrix is given by
\begin{equation}
\begin{split}
   M &= P_0(b/2)\,D_{0}^{-1}\,
       D_{1}\,P_1(a)\,D_{1}^{-1}\,D_{0}\,P_{2}(b/2)\\
\\
   &= \left(\begin{array}{cc}
    M_{11} & M_{12} \\
    M_{21} & M_{22}
 \end{array}\right) ,
 \label{matrixM}
\end{split}
\end{equation}
which may be easily calculated to obtain

\[
M_{11} = M_{22}^* = e^{\II\,b\,k_0}\,\left(
 \cos(a\,k_1)
 +\II\,\frac{k_0^2+k_1^2}{2\,k_0\,k_1}\,\sin(a\,k_1)
 \right),
\]
\[
 M_{12}=M_{21}^* = -\II\,\frac{k_0^2-k_1^2}{2\,k_0\,k_1}\,
  \sin(a\,k_1),
\]
where the asterisk indicates complex conjugate.

The vertical nulls correspond to bound states of the particle inside
the potential well, being characterized by a null reflection
coefficient, for which Eq.~\eqref{reflection:transmission} yields
\[
 \sin(a\,k_1)=\sin({\sqrt{\phi^2+\phi_V^2}})=0,
\]
having countably infinite solutions given by
\begin{equation}
 \label{cond:vertical:nulls}
 \sqrt{\phi_i^2 + \phi_V^2} = i\,\pi, \qquad
 i=1, 2, \ldots
\end{equation}

Note that the position of the vertical nulls is independent of both
the initiator length and the pre-fractal stage thanks to the
normalization of the energy~\eqref{normalizationEnergy}.

\subsection{Arc nulls} \label{ArcNulls}

The arc nulls are caused by collective bound states of the particle
in any of the $2\,N^{S-1}$ copies of the series of $\lfloor
N/2\rfloor$ potential wells wholly on either the right or left side
of each of the $N^{S-1}$ copies of the initiator at the $S$-th stage
of the pre-fractal superlattice. Since these collective bound states
depend on the distance between individual wells, they are a function
of the lacunarity parameter.

The Cayley-Hamilton theorem~\cite{LinearAlgebraBOOK} states that
every matrix $M$ satisfies the equation given by its characteristic
polynomial which for a $2\times 2$ matrix is given by
$p_M(\lambda)=\lambda^2-\tr(M)\,\lambda+\det(M)$, where $\tr(M)$ and
$\det(M)$ are the trace (sum of the diagonal elements) and
determinant, respectively, of the matrix. For the scattering matrix
of a potential well, Eq.~\eqref{matrixM}, which is an unitary
matrix, $\det(M)=|M_{11}|^2+|M_{12}|^2=1$ and $\tr(M)\in\mathbb{R}$.
Without loss of generality, $\tr(M)=2\,\cos(\theta)$, where $\theta$
is the Bloch phase, being a real number if $|\tr(M)|\le 2$ and a
complex one otherwise, being pure imaginary only if $\tr(M)>0$.
Therefore, the Cayley-Hamilton theorem yields
\[
 M^2 = 2\,\cos(\theta)\,M - I,
\]
where $I$ is the identity matrix. This equation allows the
calculation of the $n$-th power of the matrix $M$ as
\begin{equation}
 \label{powerM}
 M^n = \frac{\sin(n\,\theta)}{\sin (\theta)}\,M -
  \frac{\sin((n-1)\,\theta)}{\sin (\theta)}\,I,
\end{equation}
which can be easily proved by the induction principle, in fact, note
that
\begin{eqnarray*}
 &&
 M^{n+1} = M\,M^n =
 \frac{\sin(n\,\theta)}{\sin (\theta)}\,(2\,\cos(\theta))\,M
 -\frac{\sin(n\,\theta)}{\sin (\theta)}
 \\ && \phantom{M^{n+1} = M\,M^n = 11}
 -\frac{\sin((n-1)\,\theta)}{\sin (\theta)}\,M,
\end{eqnarray*}
and use the formula $2\,\cos(\theta)\,\sin(n\,\theta) =
\sin((n+1)\,\theta)
 +\sin((n-1)\,\theta)$.
The off-diagonal element $(M^n)_{21}$ of Eq.~\eqref{powerM} is
\[
 (M^n)_{21} = \frac{\sin(n\,\theta)}{\sin (\theta)}\,M_{21},
\]
hence the reflection coefficient is null for
\begin{equation}
 \label{sinNtheta}
 \sin(n\,\theta)=0, \qquad \sin(\theta)\ne 0,
\end{equation}
whose countably infinite solutions are the set of Bloch phases given
by
\begin{equation}
 \label{ndeij}
 n\,\theta_{ij}=(n\,i + j)\,\pi, \quad
 i=0,1,2,\ldots, \quad
 j=1,2,\ldots,n-1,
\end{equation}
where $n=\lfloor N/2\rfloor$ in our case. Note that, in the
right-hand side of this expression, $\pi$ appears multiplied by all
the natural numbers except the multiples of $n$, for which
${\sin(n\,\theta)}/{\sin (\theta)}=n$.

The Bloch phase may be calculated from Eq.~\eqref{matrixM} which
yields
\begin{eqnarray*}
 &&
 \tr(M)= 2\,
   \cos(b\,k_0)\,\cos(a\,k_1)
 \\ && \phantom{\tr(M)= 22222}
   - \frac{k_0^2+k_1^2}{k_0\,k_1}\,
     \sin(b\,k_0)\,\sin(a\,k_1)
 \\ && \phantom{\tr(M)}
  = \frac{(k_0+k_1)^2}{2\,k_0\,k_1}\,\cos(b\,k_0+a\,k_1)
 \\ && \phantom{\tr(M)= 22222}
   - \frac{(k_0-k_1)^2}{2\,k_0\,k_1}\,\cos(b\,k_0-a\,k_1).
\end{eqnarray*}
Equation~\eqref{normalizationEnergy} yields $a\,k_0=\phi$ and
\[
 a\,k_1= \sqrt{\phi^2+\phi_V^2} = \phi +
  \frac{1}{2}\,\left(\frac{\phi_V}{\phi}\right)^2
  + \OL{\left(\frac{\phi_V}{\phi}\right)^4},
\]
so for $\phi\gg\phi_V$, $k_0\approx k_1$ and
\begin{equation} \label{traceOfM}
\begin{split}
 \tr(M)&= 2\,\cos(\theta) \\
   &=2\,
   \cos(a\,k_1 + b\,k_0) + \OL{(k_1-k_0)^2}.
\end{split}
\end{equation}

Therefore, from Eq.~\eqref{ndeij}, the normalized energy $\phi_{ij}$
for the arc nulls may be calculated as function of the lacunarity as
\begin{eqnarray}
 \label{cond:arc:nulls}
 &&
 \sqrt{\phi_{ij}^2+\phi_V^2} +
 \left(\frac{\varepsilon}{\gamma}\right)^S\phi_{ij}
   = \left( i + \frac{j}{\lfloor N/2 \rfloor}\right)\pi,
 \nonumber
 \\ &&
 i=0, 1, 2, \ldots, \quad j=1, 2, \ldots, \lfloor N/2 \rfloor-1.
\end{eqnarray}

Note also that
\[
 \sin(n\,\theta)=\sin(n\,\arccos(\tr(M)/2))
\]
\[
=\frac{\sqrt{4-\tr(M)^2}}{2}\,U_n(\tr(M)/2),
\]
where $U_n(x)$ is the $n$-th Chebyshev polynomial of the second
kind~\cite{SprungAJP}. The null value of this expression at the arc
nulls is straightforwardly checked since
\[
 \tr(M)= 2\,\cos(\theta_{ij})=2\,(-1)^j,
\]
where Eq.~\eqref{ndeij} has been used.

\subsection{Striation nulls}  \label{StriationNulls}

The striation nulls are caused by the quantum interference between
the $\lfloor N/2\rfloor$ potential wells on the right with those on
the left side in each of the $N^{S-1}$ copies of the initiator at
the $S$-th stage of the pre-fractal superlattice. This interaction
is exact for even $N$ and only approximate for odd $N$ due to the
presence of the potential well at the center of each copy of the
initiator. Obviously, they also are a function of the lacunarity
parameter.

Let us start with the even $N$ case. Let $c$ be the distance between
the rightmost potential well of the $\lfloor N/2\rfloor$ wells on
the left and the leftmost one of those at the right in every copy of
the initiator at the $S$-th stage, yielding
$c=L\,(1-N\,\gamma^S-(N-2)\,\varepsilon^S)$. The scattering matrix
for this problem is $\tilde{M}^2$ where
\[
 \tilde{M} = P_0(c/2)\,M^{\lfloor N/2\rfloor}\,P_0(c/2).
\]

Equations~\eqref{sinNtheta} and~\eqref{ndeij} for $n=2$ yield
\[
 2\,\tilde{\theta}_i = (2\,i+1)\,\pi, \qquad i=0,1,2,\ldots,
\]
where $\tr(\tilde{M})=2\,\cos(\tilde{\theta})$, which may be easily
calculated from
\[
 \tr(\tilde{M})=\tr(M^{\lfloor N/2\rfloor})\,\tr(P_0(c)).
\]
Introducing the trace operator into Eq.~\eqref{powerM} yields
\[
 \tr(M^n) = \frac{\sin(n\,\theta)}{\sin (\theta)}\,\tr(M) -
  2\,\frac{\sin((n-1)\,\theta)}{\sin (\theta)}
  = 2\,\cos(n\,\theta),
\]
hence, using Eq.~\eqref{traceOfM}, for $\phi\gg\phi_V$,
\[
 \tr(\tilde{M})=2\, \cos(n\,(a\,k_1 + b\,k_0))\,2\,\cos(c\,k_0) +
 \OL{(k_1-k_0)^2},
\]
yielding $\tilde{\theta}\approx n\,a\,k_1 + (n\,b+c)\,k_0$, where
$n= \lfloor N/2\rfloor$,

resulting in
\begin{eqnarray}
 \label{cond:striation:nulls:even}
 &&
   \lfloor N/2 \rfloor \,\sqrt{\phi_i^2+\phi_V^2}+
\nonumber \\
&&
 + \left( \gamma^{-S}-N- \left( N-\lfloor N/2 \rfloor - 1\right)
   \,\left(\frac{\varepsilon}{\gamma}\right)^{S}
   \right)\,
       \phi_i=
 \nonumber \\ && \qquad
   = ( 2\,i + 1)\,\frac{\pi}{2}, \qquad
 i=0, 1, 2, \ldots,
\end{eqnarray}
for the striation nulls for even $N$.

For the odd $N$ case, the central potential well between both series
of $\lfloor N/2\rfloor$ wells makes their mutual interference not
exact. However, an approximate formula may be obtained as easily as
before. Let $c=L\,(1-N\,\gamma^S-N\,\varepsilon^S)$ such that
$c/2+b$ be the distance between the central potential well and that
in the right (or left) in every copy of the initiator at the $S$-th
stage. The corresponding scattering matrix is $\hat{M}^2$ where
\[
 \hat{M} = P_1(a/2)\,D_1^{-1}\,D_0\,P_0(c/2)\,
           M^{\lfloor N/2\rfloor}\,\cdot
\]
\[
          \cdot P_0(c/2)\,D_0^{-1}\,D_1\,P_1(a/2),
\]
where the central potential well has been divided in two parts, one
located completely to the left and the other one completely to the
right. By exactly the same arguments as before, omitted here for
brevity, $\tr(\hat{M})=2\,\cos(\hat{\theta})$, where
\[
 \hat{\theta}\approx \left( N -\lfloor N/2\rfloor \right)\,(a\,k_1 + b\,k_0)
   + c\,k_0,
\]
resulting in
\begin{eqnarray}
 \label{cond:striation:nulls:odd}
 &&
   \left( N - \lfloor N/2 \rfloor \right)\,\sqrt{\phi_i^2+\phi_V^2}+
\nonumber \\
&&
 + \left( \gamma^{-S}-N- \left( \lfloor N/2 \rfloor - 1\right)
   \,\left(\frac{\varepsilon}{\gamma}\right)^{S}
   \right)\,
       \phi_i
 \nonumber \\ && \qquad
   = ( 2\,i + 1)\,\frac{\pi}{2}, \qquad
 i=0, 1, 2, \ldots,
\end{eqnarray}
as an approximation for the striation nulls for odd $N$.

Note that Eq.~\eqref{cond:striation:nulls:even} is a particular case
of Eq.~\eqref{cond:striation:nulls:odd}, hence in
Ref.~\cite{JaggardJOSA} only the last one is shown for both even and
odd $N$.

\section{Presentation of results}
\label{Results}

\begin{figure}
 \begin{center}
  \includegraphics[width=7cm]{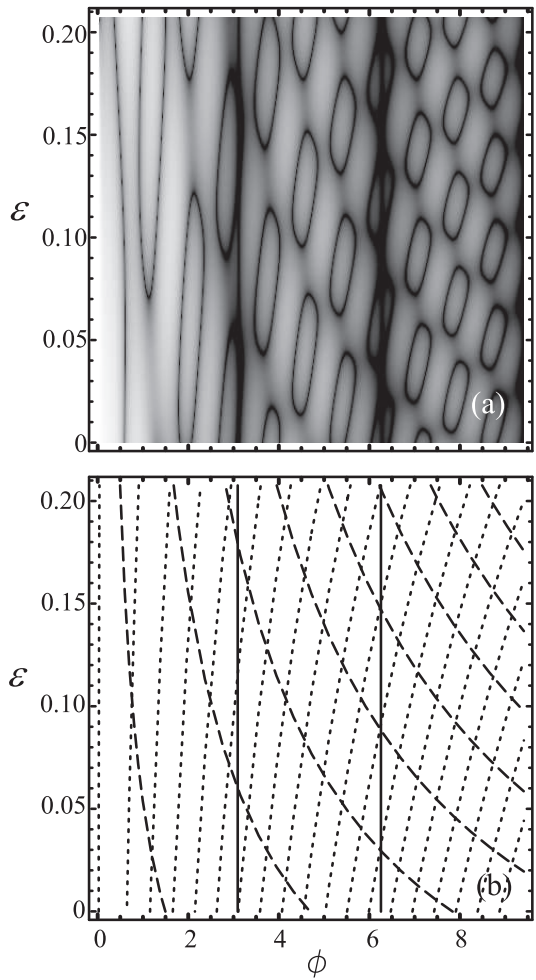}
 \end{center}
  \caption{
  Scattering reflection coefficient (a) for the
  polyadic Cantor pre-fractal potentials with $S=1$, $N=5$,
  $\gamma=1/7$, and $\phi_V=1/2$, and the
  analytical curves (b) for the corresponding vertical (continuous line),
  arc (dashed line), and striation (dotted line) nulls.}
 \label{figuraRforPolyadicCantorNcinco}
\end{figure}

\begin{figure}
 \begin{center}
  \includegraphics[width=7cm]{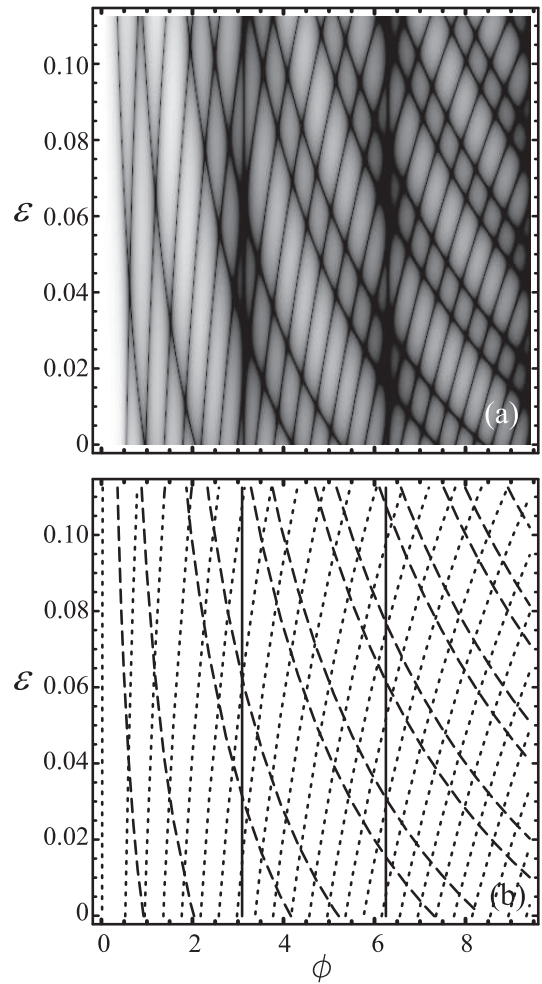}
 \end{center}
  \caption{Scattering reflection coefficient (a) for the
  polyadic Cantor pre-fractal potentials with $S=1$, $N=6$,
  $\gamma=1/7$, and $\phi_V=1/2$, and the
  analytical curves (b) for the corresponding vertical (continuous line),
  arc (dashed line), and striation (dotted line) nulls.}
 \label{figuraRforPolyadicCantorNseis}
\end{figure}

Twist plots are the gray-scale representation of the scattering
reflection coefficient (in decibels, i.e., $10\,\log_{10}R$) as a
function of the normalized energy ($\phi$) and the lacunarity
parameter ($\varepsilon$). Here, a linear gray scale was used, from
black for null values to white for the maximum value equal to unity.
Figures~\ref{figuraRforPolyadicCantorNcinco}~(a)
and~\ref{figuraRforPolyadicCantorNseis}~(a) show twist plots for,
respectively, pentadic ($N=5$) and hexadic ($N=6$) Cantor fractal
potential both with $S=1$, $\gamma=1/7$, and $\phi_V=1/2$. The
horizontal interval covers the normalized energy $\phi$ from $0$ to
$\sqrt{(3\,\pi)^2-\phi_V^2}\approx 9.41$, the energy of the third
vertical null line. These figures have been obtained by a
straightforward numerical implementation of the transfer matrix
method.

The most noticeable fact in
Figs.~\ref{figuraRforPolyadicCantorNcinco}~(a)
and~\ref{figuraRforPolyadicCantorNseis}~(a) is the null structure
(black lines and curves) corresponding to energies at which the
particle transparently tunnels through the fractal potential wells.
The strongest (darkest) among these curves are the vertical nulls.
The arc nulls, which are concave upward, curving from the upper left
to the lower right of the plots, appear in pairs for $N=6$ (see
Fig.~\ref{figuraRforPolyadicCantorNseis}~(a)), as expected from
Eq.~\eqref{cond:arc:nulls}. The striation nulls, which are concave
downward, running from the lower left to the upper right of each
twist plot, form the finer structure of the twist plots.

The main difference between
Figs.~\ref{figuraRforPolyadicCantorNcinco}~(a)
and~\ref{figuraRforPolyadicCantorNseis}~(a) is observed for $N=5$
and, in general, for odd $N$, because in such a case the tunnelling
is not perfectly transparent at the whole length of the arc nulls,
but instead appear dashed, alternatively darker and lighter after
they cross the striation nulls. The striation nulls also appear
lighter when they cross the arc nulls, resulting in a leopard-like
spot pigmentation pattern. This result is not unexpected, since both
the arc and the striation nulls are not truly null for odd $N$ due
to the single quantum well between each pair of $\lfloor N/2\rfloor$
potential wells, which introduces non-periodicity in the scattering
problem resulting in a (small) non-null reflection coefficient.

The nulls calculated by Eqs.~\eqref{cond:vertical:nulls},
\eqref{cond:arc:nulls}, and~\eqref{cond:striation:nulls:odd}, are
shown in Figs.~\ref{figuraRforPolyadicCantorNcinco}~(b)
and~\ref{figuraRforPolyadicCantorNseis}~(b). These figures clearly
show the good accuracy of the analytical approximations which
reproduce the main features of the corresponding twist plots. For
even $N$ the accuracy is surprisingly good. For odd $N$, as
expected, both arc and striation nulls are less accurate. Note also
that the accuracy degrades for small energy as expected due to the
lack of validity of the approximations used in
Sections~\ref{ArcNulls} and~\ref{StriationNulls} in such a case.

\section{Conclusions}
\label{Conclusions}

Analytical expressions for the calculation of the particle energy
for transparent tunnelling in polyadic Cantor pre-fractal potentials
as a function of the lacunarity parameter have been obtained from
first principles by using the transfer matrix method. The comparison
with the results obtained by a numerical implementation of the
transfer matrix method shows the good accuracy of these expressions
for $N$-adic Cantor pre-fractals with even $N$ and their reasonable
accuracy for odd $N$. The reasons for this difference in accuracy
have also been presented in detail.

The analytical results obtained in this paper can be easily
incorporated as a complement in computer laboratories for
undergraduate quantum mechanics or solid-state physics courses
currently using the transfer matrix method only as a numerical
tool~\cite{VillatoroEJP,VillatoroAJP}. In fact, the possibility of
approximately calculating the geometrically complex figures observed
in the results provides the student with an excellent example of the
combination of theory and numerical experiments. Thus, different
aspects of the model may be assigned to different students or groups
of students.  In fact, stimulating discussions among the students
follow naturally from this kind of approach.

Finally, since the transfer matrix methods may be adapted by physics
analogy to the propagation of waves in general one-dimensional
quasiperiodic media~\cite{GriffithsAJP}, the present results may be
also applied to, for example, acoustics, optics, or vibrating
strings. In fact, the effect of the lacunarity in the scattering on
generalized Cantor media was first studied in
optics~\cite{JaggardJOSA,MonsoriuOE}, where the additional degree of
freedom introduced by the lacunarity was used with success to obtain
new physical characteristics. A detailed study of these
applications, outside the scope of this paper, may also be an
estimulating topic for further study.

\begin{acknowledgements}
This work has been supported by the Ministerio de Educaci\'on y
Ciencia (grant FIS2005-01189), Spain. We also
acknowledge the financial support from the Universidad
Polit\'ecnica de Valencia (Vicerrectorado de Innovaci\'on y Desarrollo,
Programa de Incentivo a la Investigaci\'on 2005), Spain.
\end{acknowledgements}

\newcommand{\bookref}[5]{#1, {\em {#2}}, (#3, #4, #5).}

\newcommand{\paperref}[6]{#1, ``#2," {\em {#3}} {\bf #4}, #5 (#6).}
\newcommand{\paperrefno}[7]{#1, ``#2," {\em {#3}} {\bf #4} (#5), #6 (#7).}

\end{document}